# Heteroepitaxial Thin-Film Growth of a Ternary Nitride Semiconductor CaZn$_2$N$_2$

Masatake Tsuji,[1] Kota Hanzawa,[1] Hiroyuki Kinjo,[1] Hidenori Hiramatsu,[1,2,*] and Hideo Hosono [2]

1: Laboratory for Materials and Structures, Institute of Innovative Research, Tokyo Institute of Technology, Mailbox R3-3, 4259 Nagatsuta-cho, Midori-ku, Yokohama 226-8503, Japan

2: Materials Research Center for Element Strategy, Tokyo Institute of Technology, Mailbox SE-1, 4259 Nagatsuta-cho, Midori-ku, Yokohama 226-8503, Japan

*) E-mail: h-hirama@mces.titech.ac.jp

**Abstract**

Zinc-based nitride $CaZn_2N_2$ films grown by molecular beam epitaxy (MBE) with a plasma-assisted active nitrogen-radical source are promising candidates of next-generation semiconductors for light-emitting diodes and solar cells. This nitride compound has previously only been synthesized in a bulk form by ultrahigh-pressure synthesis at 5 GPa. Three key factors have been found to enable heteroepitaxial film growth: (i) precise tuning of the individual flux rates of Ca and Zn, (ii) the use of GaN template layers on sapphire *c*-plane as substrates, and (iii) the application of MBE with an active N-radical source. Because other attempts at physical vapor deposition and thermal annealing processes have not produced $CaZn_2N_2$ films of any phase, this rf-plasma-assisted MBE technique represents a promising way to stabilize $CaZn_2N_2$ epitaxial films. The estimated optical band gap is ~1.9 eV, which is consistent with the value obtained from bulk samples. By unintentional carrier doping, n- and p-type electronic conductions are attained with low carrier densities of the order of $10^{13}$ $cm^{-3}$. These features represent clear advantages when compared with Zn-based oxide semiconductors, which usually have much higher carrier densities irrespective of their intentionally undoped state. The carrier mobilities at room temperature are 4.3 $cm^2/(V·s)$ for electrons and 0.3 $cm^2/(V·s)$ for hole carriers, which indicates that transport properties are limited by grain boundary scattering, mainly because of the low-temperature growth at 250 °C, which realizes a high nitrogen chemical potential.

**Introduction**

New functional materials, composed only of earth abundant elements, are required to realize sustainable and environmentally friendly devices. In particular, energy storage, energy conversion, and light emission are key required functionalities, as represented by light-emitting diodes, laser diodes, and solar cells, based on semiconductor devices. Currently, light-emitting diodes and laser diodes are produced from III–V nitride, phosphide, and arsenide semiconductors, such as (Al, Ga, In)N and (Al, Ga, In)(P, As). However, despite rapid technological improvements, light-emitting semiconductors based on III–V group compounds face a serious issue, known as the green gap problem: the emission quantum efficiency decreases to ~20% over a particular green wavelength region compared with the efficiencies in the blue and red regions of ~80% [1]. This effect arises mainly because of the large in-plane lattice mismatch between In-based III–V group compounds and single-crystal substrates and the thermal instability of In-based III–V nitrides. Thus, no highly efficient green-light-emitting semiconductor material has been realized for next-generation optoelectronic devices that require high performance in terms of light brightness, quantum efficiency, and color accuracy (i.e., emission bandwidth). Notably, $Cu(In, Ga)Se_2$ (CIGS) devices are based on the materials that give highly efficient solar cells. However, rare elements such as In and toxic elements such as As and Se are primary elements used in the above functional practical optoelectronic devices. The required optical band gaps for these semiconductors are in the range of 2.5–1.8 eV.

We have recently proposed ternary Zn-based nitride semiconductors as a candidate for next-generation functional optoelectronic semiconductors because of their environmentally benign nature. These materials remain relatively unexplored to date compared with other materials systems such as oxides, chalcogenides, and pnictides. Among them, we focused on $CaZn_2N_2$ [2] and its alloy system $Ca(Mg_{1-x}Zn_x)_2N_2$ ($x$ = 0–1) [3] because they have small electron and hole effective masses and a direct a transition-type energy band structure, and their optical band gaps can be continuously tuned from 3.2 to 1.8 eV [3] by traditional semiconductor alloying techniques similar to those used for III–V nitrides and phosphides. Band-to-band photoluminescence is observed from these materials at room temperature in the ultraviolet–red region depending on $x$; e.g., the $x$ = 0.50 sample exhibits intense green emission with a peak at 2.45 eV (506 nm) [3]. In addition, 2% Na doping at the Ca site of $CaZn_2N_2$ converts its highly resistive state to a p-type conducting state. These features of Ca(Mg$_{1-x}$Zn$_x$)$_2$N$_2$ clearly meet the demands of III–V group nitride and arsenide/phosphide light-emitting semiconductors and the optical absorption layers of solar cells.

However, Ca(Mg$_{1-x}$Zn$_x$)$_2$N$_2$ ($x$ = 0–1) samples are polycrystalline bulk materials [2, 3], synthesized at an ultrahigh pressure of 5 GPa and have not previously been made in a thin-film form suitable for device applications. The development of appropriate thin-film growth techniques is therefore necessary to realize next-generation semiconductor devices based on these materials. This would be the most important for future application such as photovoltaic and light-emitting electronic devices. In this study, we work toward thin-film growth of one (i.e., $x$ = 1) of the end-members of Ca(Mg$_{1-x}$Zn$_x$)$_2$N$_2$ ($x$ = 0–1) and succeed in it by a molecular beam epitaxy technique with a radio-frequency-generated active N-radical source (RF-MBE). This is the first demonstration of the fabrication of CaZn$_2$N$_2$ epitaxial thin films.

**Experimental procedures**

Heteroepitaxial CaZn$_2$N$_2$ thin films were grown by RF-MBE on substrates, heated by an electrical resistance heater at substrate temperatures ($T_s$) between 250 and 300 °C. We used ~5 μm thick GaN (0001) template layers on $\alpha$-Al$_2$O$_3$ (0001) single crystals, Y-stabilized ZrO$_2$ (111) single crystals (YSZ), and silica glass as substrates for the growth. The YSZ was thermally annealed at 1350 °C in air before use; whereas, the other substrates were used as received. During the thin-film growth, the internal nitrogen gas pressure of the MBE growth chamber, which had a base pressure of $2\times10^{-8}$ Pa, was set to be $1\times10^{-2}$, $7\times10^{-3}$, and $4\times10^{-3}$ Pa through the introduction of an active N radical source generated by rf power of >300 W (i.e., bright mode), where distance from the substrate and the rf generator was set to be 120 mm. The chemical composition of the resulting CaZn$_2$N$_2$ film was tuned by individually changing the beam fluxes of Zn ($P_{Zn}$) and Ca ($P_{Ca}$). The former ranged from $7.5\times10^{-5}$ to $2.9\times10^{-4}$ Pa and the latter from $1.4\times10^{-5}$ to $1.2\times10^{-4}$ Pa, both evaporated from Knudsen cells at cell temperatures of 215–300 °C for Zn and 460–540 °C for Ca, which were equipped at 230 mm far from the substrate.

Crystalline phases of the fabricated films were characterized by X-ray diffraction (XRD) with incident Cu K$\alpha_1$ and K$\alpha_2$ radiation (averaged $\lambda$ = 1.541 Å) with Bragg–Brentano geometry. High-resolution XRD (HR-XRD) with a monochromated Cu K$\alpha_1$ source ($\lambda$ = 1.540562 Å) was used to evaluate the structure parameters such as lattice parameters. The crystallite sizes for out-of-plane and in-plane were estimated from the Sherrer's equation (crystallite size $D = K\lambda / (\Delta\theta \cos\theta)$, where $K$, $\lambda$, and $\Delta\theta$ denote Sherrer constant ($K = 0.94$), the wavelength of X-ray, and full width at half-maxima of CaZn$_2$N$_2$ 0001, 0003 and 11$\bar{2}$0 diffraction, respectively). The crystallite orientation and crystallinity were determined from measurements of the out-of-plane rocking curve at CaZn$_2$N$_2$ 0001 diffraction. The thicknesses ($t$) of the films were measured precisely by X-ray reflectivity when $t < 100$ nm and roughly with a stylus profiler for $t > 100$ nm. The surface roughness of the films was characterized from the surface morphology observed with an atomic force microscope (AFM). Chemical composition analysis was performed by electron-probe microanalysis (EPMA) or field-emission scanning Auger electron spectroscopy (FE-AES). In FE-AES measurements, the incident electron beam was weakly accelerated at 10 kV and 10 nA to suppress electrical charge-up of the specimens. The depth profiles were established through Ar sputtering of the films. Optical properties of the obtained films were characterized by transmittance ($T$) and normal reflectance ($R$) measurements in the visible−near-infrared wavelength region ranging from 200 to 2500 nm at room temperature. The absorption coefficient ($\alpha$) was calculated with the equation; $\alpha = \ln[(1-R)/T]/t$. We used $(\alpha h\nu)^2$ vs. $h\nu$ plots, where $h$ and $\nu$ are respectively the Planck constant and photon frequency, to determine the direct band gap. The electronic transport properties were evaluated at room temperature by Hall-effect measurements using a van der Pauw configuration under an ac magnetic field of 0.3 T for CaZn$_2$N$_2$ on silica glass and 0.4 T on YSZ, where the detection limit of Hall mobility is as low as $1\times10^{-3}$ cm$^2$/(V·s).

Results and discussion

Prior to heteroepitaxial thin-film growth of CaZn$_2$N$_2$ by RF-MBE, we attempted growth by *in situ* pulsed laser deposition (PLD) and postdeposition thermal annealing via the reactive solid-phase epitaxy (R-SPE) technique [4], which is an effective film-growth technique for heteroepitaxial growth of complex compounds with high vapor pressure elements such as Zn, Cu, and related chalcogenides [5–8]. However, we could not obtain a CaZn$_2$N$_2$ phase by these approaches. The details are explained in the Supporting Information (Text S1). These results clarified that the equilibrium process occurring through thermal annealing is not suitable for growing CaZn$_2$N$_2$. Additionally, one nonequilibrium processes, *in situ* PLD, could not also achieve a stoichiometric chemical composition because of an excessively high vapor pressure of Zn.

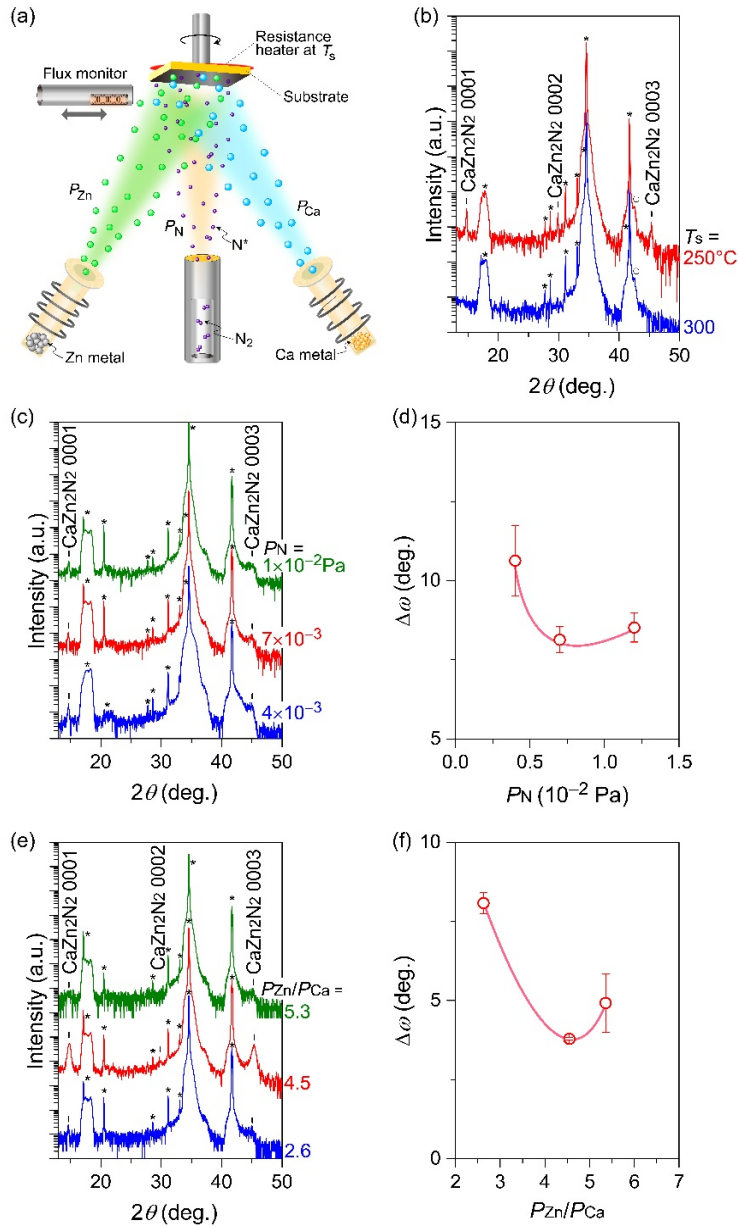

**Figure 1.** Optimization of growth conditions of $CaZn_2N_2$ films on GaN template layers by RF-MBE. (a) Illustration of RF-MBE growth. (b) Substrate temperature ($T_s$) dependence of XRD patterns. Circles and asterisks denote diffraction features originating from impurities in the Ca–Zn alloy and substrate, respectively. (c) XRD patterns of $CaZn_2N_2$ thin films grown at different

pressures ($P_N$) of the rf-generated active N-radical source at 350 W. (d) Relationship between full width at half-maxima ($\Delta\omega$) of $CaZn_2N_2$ and the $P_N$. The value of $\Delta\omega$ was estimated from the rocking curves at the $CaZn_2N_2$ 0001 diffraction in **Figure S5a**. Bars at each red circle denote the standard errors. (e) XRD patterns of $CaZn_2N_2$ grown at different beam flux ratios of Zn ($P_{Zn}$) and Ca ($P_{Ca}$). (f) $\Delta\omega$ of $CaZn_2N_2$ deposited with different $P_{Zn}/P_{Ca}$. The diffraction patterns of rocking curves and the fitting results are summarized in **Figure S5b**.

The above preliminary experimental results confirmed that conventional and traditional film-growth techniques were inappropriate for film growth of $CaZn_2N_2$. These results suggest that fabrication of $CaZn_2N_2$ epitaxial thin films requires a nonequilibrium thin-film growth process with an active N-radical source to achieve a sufficiently high N-chemical potential and effectively stabilize the $CaZn_2N_2$ phase, where the chemical composition can precisely be tuned. Hence, we selected RF-MBE with a bright-mode N plasma (i.e., N* radicals rather than dark mode $N_2$*) (see **Figures 1a, S3b, and S3c**). The results shown in **Figure S3** indicate that $CaZn_2N_2$ was effectively stabilized under a N plasma with a bright mode rf generated at >300 W rather than by the dark mode at <300 W. **Figure 1** summarizes the optimization of growth conditions of $CaZn_2N_2$ by RF-MBE. We first optimized $T_s$ in **Figure 1b** on GaN (0001) templates formed on $\alpha$-$Al_2O_3$ (0001) at a vapor pressure ratio of $P_{Zn}/P_{Ca} = 2.4$ at an internal active N radical gas pressure ($P_N$) of $7\times10^{-3}$ Pa. In this optimization, we used a $T_s$ value lower than 320 °C because of the decomposition temperature of $CaZn_2N_2$ (320 °C) determined from thermal desorption spectroscopy of $CaZn_2N_2$ powder (see **Figure S4**), at which point major constituents such as Zn started to evaporate. At $T_s = 300$ °C, only an impurity phase, characterized as a Ca–Zn binary alloy, such as $CaZn_5$ was segregated, indicating a lack of N chemical potential for nitridization at this temperature. Thus, we decreased $T_s$ and consequently succeeded in stabilizing the $CaZn_2N_2$ phase at a lower $T_s = 250$ °C. We note that because our research target is heteroepitaxial growth of $CaZn_2N_2$, thermal energy is required to promote epitaxial growth. Hence, we concluded that 250 °C is the optimum $T_s$ to grow epitaxial $CaZn_2N_2$ films. Next, we investigated the effects of $P_N$ on the growth at $T_s = 250$ °C and $P_{Zn}/P_{Ca} = 2.4$.

The XRD results in **Figure 1c** indicate that a single phase of $CaZn_2N_2$ was obtained under the examined $P_N$ ranging from $4\times10^{-3}$ to $1\times10^{-2}$ Pa. However, rocking-curve measurements at the $CaZn_2N_2$ 0001 diffraction peak (data summarized in **Figure S5a**) suggested that the crystallinity ($\Delta\omega$) exhibited an inverted-bell-type behavior with respect to $P_N$, as shown in **Figure 1d**; i.e., $\Delta\omega$ decreased as $P_N$ increased and then started to increase. The $\Delta\omega$ value of the $CaZn_2N_2$ grown at $7\times10^{-3}$ Pa was the narrowest ($\Delta\omega = 8.1°$) among the samples, suggesting that a $P_N$ of $4\times10^{-3}$ Pa provided a high enough chemical potential to stabilize the $CaZn_2N_2$ phase, whereas an excessively high $P_N$ prevented effective crystal growth. Hence, there is a critically optimum $P_N$ for growth of $CaZn_2N_2$ films. Finally, the atomic ratio was tuned to the optimum $T_s$ of 250 °C and $P_N$ of $7\times10^{-3}$ Pa by changing the individual beam fluxes of Zn and Ca, where $P_{Zn}$ was set to be $1\times10^{-4}$ Pa and $P_{Ca}$ was set to be $1.4\times10^{-5}$ ($P_{Zn}/P_{Ca} = 5.3$), $2.2\times10^{-5}$ (4.5), and $3.8\times10^{-5}$ Pa (2.6), corresponding to growth rates of 1.0, 1.7, and 3.8 nm/min, respectively. **Figure 1e** shows XRD patterns of the films grown at different $P_{Zn}/P_{Ca}$. All the films exhibited *c*-axis orientation of $CaZn_2N_2$. **Figure 1f** shows the relationship between the $P_{Zn}/P_{Ca}$ and the crystallinity, where diffraction patterns are summarized in **Figure S5b**. This relationship suggests that $CaZn_2N_2$ fabricated at $P_{Zn}/P_{Ca} = 4.5$ has the highest out-of-plane crystallinity ($\Delta\omega = 3.8°$); hence, the chemical composition should also be close to the true stoichiometry. We precisely characterized the atomic ratio of [Zn]/[Ca] by EPMA; however, [N]/[Ca] could not be evaluated because of the influence of N in the GaN substrate. The [Zn]/[Ca] was 2.1, which is very close to the stoichiometry of 2.0, suggesting that the chemical composition close to the required stoichiometry should have the highest crystallinity.

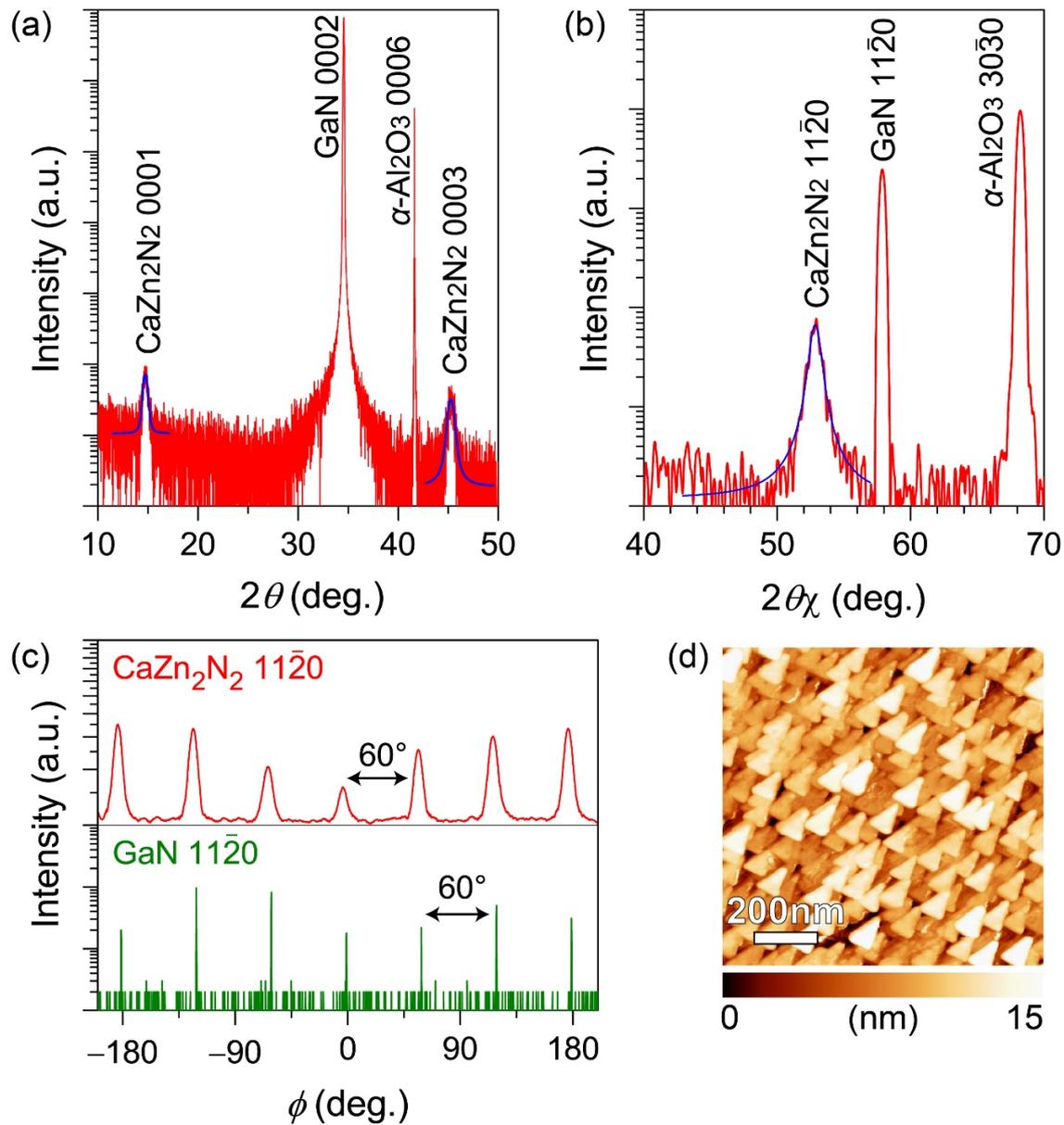

**Figure 2.** Structure and surface analyses of CaZn$_2$N$_2$ epitaxial thin film with ~50 nm thickness on GaN/α-Al$_2$O$_3$. (a–c) HR-XRD patterns of CaZn$_2$N$_2$ grown at 250 °C under $P_N = 7\times10^{-3}$ Pa with $P_{Zn}/P_{Ca} = 4.5$ for (a) out-of-plane and (b) in-plane. Blue curves are the fitting results of diffraction peaks from CaZn$_2$N$_2$. (c) ϕ scan at CaZn$_2$N$_2$ (top) and GaN (bottom) 11$\bar{2}$0 diffraction peaks. (d) Surface morphology of CaZn$_2$N$_2$ epitaxial thin films observed by AFM. The horizontal bar is an indicator of the height scale.

For the CaZn$_2$N$_2$ film grown optimally on GaN /α-Al$_2$O$_3$, we performed detailed structural analyses by HR-XRD. **Figures 2a** and **2b** show the out-of-plane and in-plane XRD patterns, respectively, in which the blue lines denote the fitting results of the CaZn$_2$N$_2$ peaks. From the fitting, we estimated the *a*- and *c*-axes lattice parameters to be 3.463 and 6.009 Å, and the crystallite size to be ~10 and ~15 nm, respectively. These results were almost the same as those of polycrystalline bulk (*a* = 3.46380 and *c* = 6.00969 Å [2]), suggesting that lattice strain was fully released, irrespective of the low thickness of ~50 nm and the large in-plane lattice mismatch between GaN (*a* = 3.19 Å) and CaZn$_2$N$_2$. **Figure 2c** shows the results of $\phi$ scans at the CaZn$_2$N$_2$ (top) and GaN 11$\bar{2}$0 (bottom) diffractions. Because clear 6-fold symmetry derived from their hexagonal structures was observed, we concluded that CaZn$_2$N$_2$ grew heteroepitaxially on GaN because of the highly active N source (i.e., bright-mode N radical gas) and precisely controlled chemical composition with the use of MBE. The epitaxial relationship was [0001] CaZn$_2$N$_2$ ∥ [0001] GaN ∥ [0001] α-Al$_2$O$_3$ for the out-of-plane and [11$\bar{2}$0] CaZn$_2$N$_2$ ∥ [11$\bar{2}$0] GaN ∥ [10$\bar{1}$0] α-Al$_2$O$_3$. To our knowledge, this is the first demonstration of the fabrication of CaZn$_2$N$_2$ epitaxial thin films. The surface morphology is shown in **Figure 2d**, we observed that triangle facet-like structures, which should reflect hexagonal structures, aligned along the same direction, supporting our conclusion of epitaxial growth. From the AFM image, we estimated the root-mean-square roughness ($R_{rms}$) to be 3.0 nm because of the island-growth mode, which resulted in large and tall grains with many related grain boundaries. From these results, we concluded that the CaZn$_2$N$_2$ epitaxial thin film is optimally grown by MBE with $P_{Zn}/P_{Ca}$ = 4.5 at $T_s$ = 250 ºC and $P_N$ = 7×10$^{-3}$ Pa.

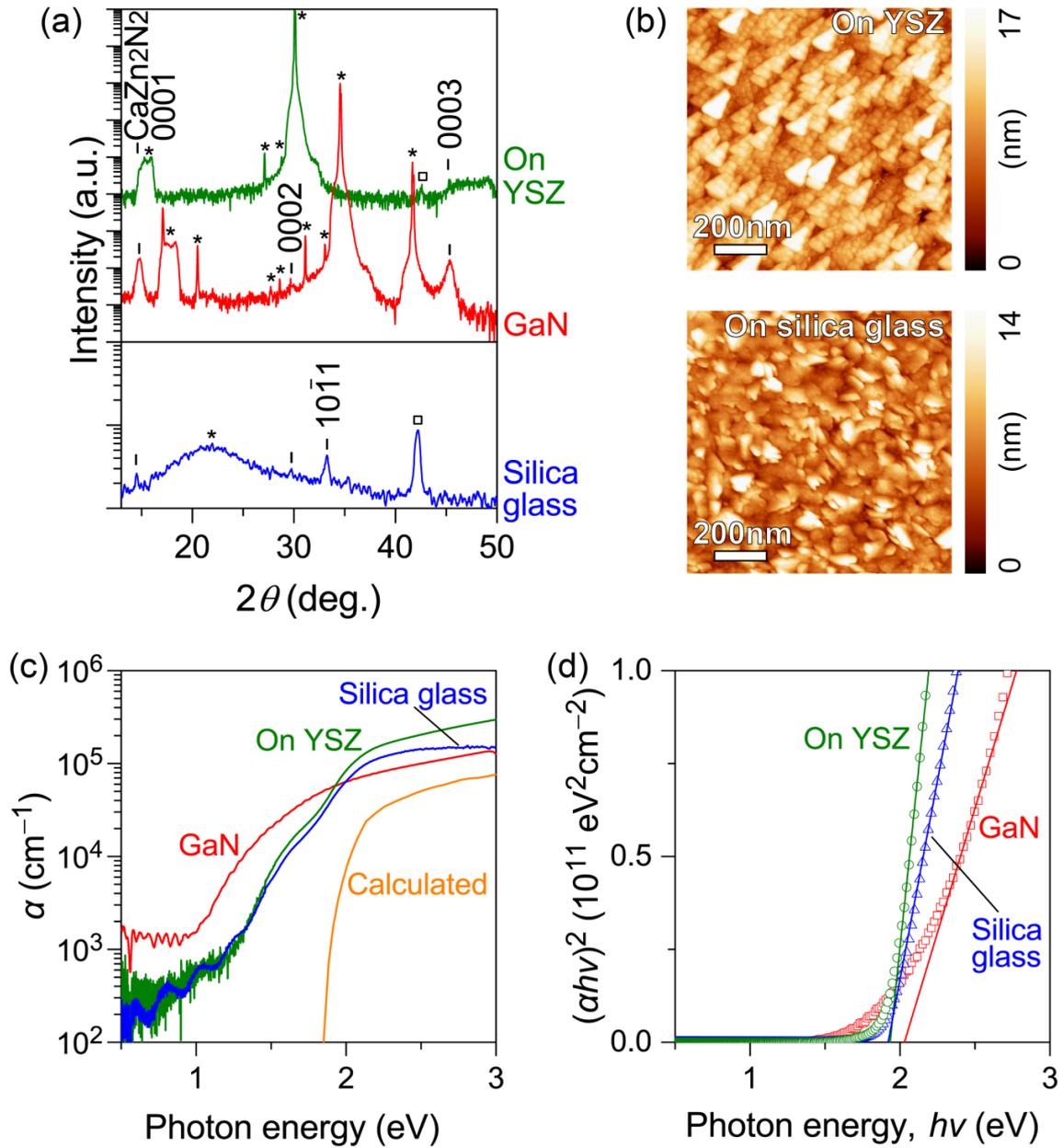

**Figure 3.** Structural and optical properties of $CaZn_2N_2$ thin films grown on YSZ, GaN, and silica glass. (a) XRD patterns of the $CaZn_2N_2$ thin films. Asterisks and squares respectively denote diffraction peaks from the substrates and an impurity phase (possibly $CaZn_5$). (b) AFM images of $CaZn_2N_2$ films fabricated on YSZ (top) and silica glass (bottom). Vertical indicators represent the height scales. (c) Absorption coefficients ($\alpha$) of $CaZn_2N_2$ deposited on the substrates and from first-principles calculation [2]. (d) $(\alpha h\nu)^2$ vs. $h\nu$ plots to evaluate the direct band gap constructed from $\alpha$ spectra in (c). The linear lines are fitting results.

Next, we examined the influence of the substrates on the structural parameters and surface morphology. **Figure 3a** shows out-of-plane XRD patterns of $CaZn_2N_2$ fabricated on silica glass (bottom), GaN (center), and YSZ (top), where $t$ was set to be ~640, ~50, and ~150 nm, and the growth conditions were independently optimized on each substrate, as for $CaZn_2N_2$ on GaN (**Figure 1)**. In the cases of growth on YSZ and GaN, we detected a clear $c$-axis orientation of $CaZn_2N_2$, whereas on silica glass, random orientations such as the $CaZn_2N_2$ $10\bar{1}1$ diffraction were observed because of its amorphous substrate characteristics. The surface of $CaZn_2N_2$ on YSZ was similar to that of GaN, with triangle grains appearing with the same surface roughness ($R_{rms}$ = 3.0 nm). This surface morphology suggested that $CaZn_2N_2$ was also epitaxially grown on YSZ, although we did not detect in-plane XRD features of $CaZn_2N_2$ on YSZ, such as $11\bar{2}0$ owing mainly to insufficient crystallinity. In contrast to the surface structures of $CaZn_2N_2$ on GaN and YSZ, that on silica glass was more disordered and flatter ($R_{rms}$ = 2.5 nm) because of the random orientation and the resultant dense structure along the $ab$-plane.

For all the films, we investigated the optical transmittance in the visible–near-infrared wavelength region at room temperature. The absorption coefficient ($\alpha$) and the corresponding $(\alpha h\nu)^2$ vs. $h\nu$ plots are shown in **Figures 3c** and **3d**. The value of $\alpha$ was as high as $1\times10^5$ cm$^{-1}$, which is comparable to that of GaAs (~$7\times10^4$ cm$^{-1}$ [9]), although the thresholds of the direct transition were not sharp. Next, we estimated the band gaps to be 1.93 eV for $CaZn_2N_2$ films grown on silica glass and YSZ (**Figure 3d**), which is consistent with that of the bulk reported in ref [2] (1.8 eV at room temperature), whereas the $CaZn_2N_2$ on GaN exhibited a larger band gap of 2.04 eV. Here, we note the large subgap state was observed, particularly in $CaZn_2N_2$ on GaN (**Figure 3c**). Thus, we concluded that the band gap of $CaZn_2N_2$ on GaN was overestimated because of the existence of the subgap states, which hide the intrinsic band edge. Owing to the large amount of subgap states, we could not observe photoluminescence originating from the band-to-band transition in any of the films, irrespective of the clear observations of band-to-band photoluminescence in polycrystalline bulk samples [2, 3]. The origin of the subgap state will be discussed from chemical compositional analysis with the use of FE-AES.

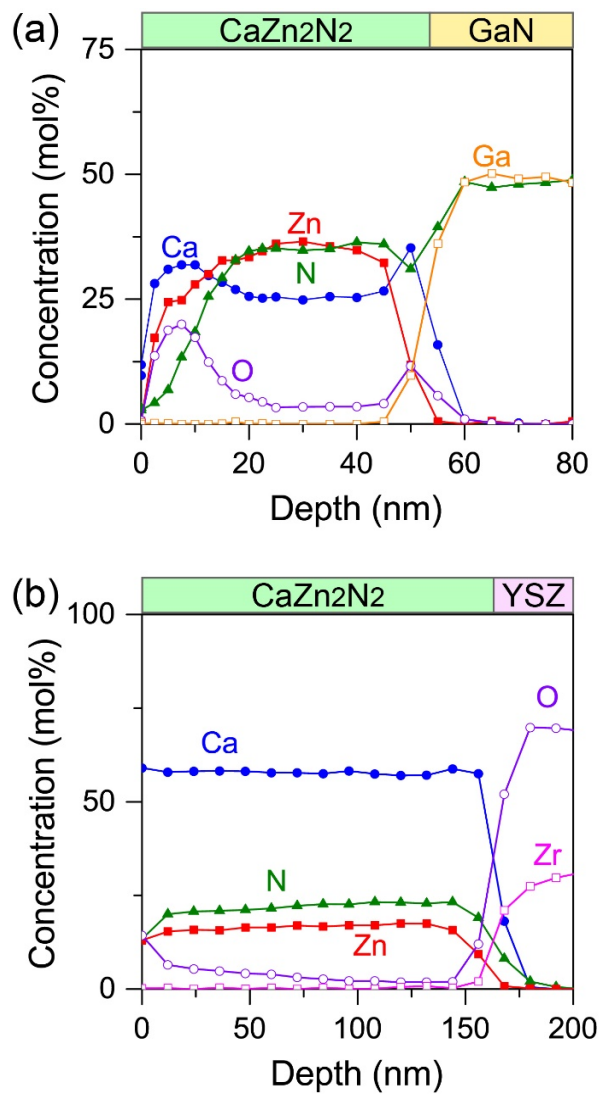

**Figure 4.** Chemical composition analyses by FE-AES for $CaZn_2N_2$ on (a) GaN and (b) YSZ.

**Table 1.** Summary of Hall Effect Measurements of $CaZn_2N_2$ Films at Room Temperature

| Substrate | Carrier type | Electrical resistivity, $\rho$ ($\Omega\cdot$cm) | Carrier density, $n$ (cm$^{-3}$) | Mobility, $\mu$ [cm$^2$/(V·s)] |
|---|---|---|---|---|
| YSZ | p | $1\times10^6$ | $2\times10^{13}$ [$(7-70)\times10^{12}$] | 0.3 (0.1 – 1) |
| Silica glass | n | $2\times10^4$ | $6\times10^{13}$ [$(5-10)\times10^{13}$] | 4 (3 – 6) |

Next, we measured the electronic transport properties of $CaZn_2N_2$ deposited on YSZ and silica glass from Hall-effect measurements with the use of the van der Pauw configuration (**Table 1**). Here, we note that we could not obtain reliable results for $CaZn_2N_2$ on GaN because the substrate also features non-negligible n-type conduction, where $O^{2-}$ incorporates into $N^{3-}$ sites in $CaZn_2N_2$ and acts as donors [10, 11]. We clarified that the dominant carrier types were p-type (holes) and n-type (electrons) for $CaZn_2N_2$ on YSZ and silica glass, respectively. The estimated bulk carrier densities ($n$) of the thin films on YSZ and silica glass were as low as $2\times10^{13}$ and $6\times10^{13}$ cm$^{-3}$. This result is notable when compared with the $n$ values of Zn-based oxide epitaxial films such as ZnO, which feature $n$ as high as $1\times10^{16}-10^{18}$ cm$^{-3}$ [12]. Hence, $CaZn_2N_2$ shows potential for controlling carriers over a wider range from intrinsic (i.e., close to insulating), semiconducting, through to degenerate states by impurity doping. The low $n$ value is reflected in the high bulk resistivities ($\rho$) of $1\times10^6$ and $2\times10^4$ $\Omega\cdot$cm. From these values, we calculated the carrier mobilities. Irrespective of the small effective masses of both electrons (0.17 $m_0$) and holes (0.91 $m_0$) [2], the mobilities were as low as 4.3 for electrons and 0.3 cm$^2$/(V·s) for holes in $CaZn_2N_2$. We attribute the low mobilities to the poor crystallinity of the films. Although the different carrier types lead to different mobilities, i.e., the mobility of electrons should be roughly 5 times as high as that of holes in $CaZn_2N_2$, this difference cannot explain the 1 order of magnitude higher mobility on silica glass than that on YSZ. Therefore, we concluded that the difference in mobilities is likely related to the surface structure (i.e., grain boundary scattering) rather than the difference between the effective masses. In the case of the mobility on YSZ, large grains of $CaZn_2N_2$ grew independently, which contributed to poor overlap between the wave functions and large potential barriers between the grains.

The n-type carriers in $CaZn_2N_2$ on silica glass should be supplied through generation of N vacancies, as has been predicted by first-principles calculation under N-deficient conditions (i.e., a relatively deficient N-chemical potential during film growth) [2], whereas the origin of p-type carriers is unclear from the calculations. To explore the origin of p-type carrier conduction in $CaZn_2N_2$ on YSZ and the large subgap states in the films grown on GaN, we performed chemical compositional analyses with FE-AES along the depth direction. **Figures 4a** and **4b** show the relationship between the atomic concentration estimated from the signal intensity of Auger electrons and detection depth in $CaZn_2N_2$ films on GaN and YSZ with respect to the constituents of $CaZn_2N_2$ and the substrates. In the $CaZn_2N_2$ film on GaN, a large amount of incorporated O was detected over the whole film and the interface between the film and GaN, despite the non-oxide nature of the GaN substrate. Thus, we concluded that O replacing N sites formed large subgap states, similar to those of O-doped (Al, Ga)N [11]. Moreover, we observed O contamination at the surface of GaN, supporting our hypothesis for n-type conduction of GaN, although the origin of the O incorporation remains unclear. Here, we consider the atomic ratio [Zn]/[Ca] of $CaZn_2N_2$ on GaN ([Zn]/[Ca] = 2.1), as estimated from EPMA, to be very close to the formula stoichiometry. Thus, the chemical composition of $CaZn_2N_2$ probed by FE-AES reproduces that quantified by EPMA. Conversely, for $CaZn_2N_2$ on GaN, O contamination of $CaZn_2N_2$ on YSZ was less pronounced, irrespective of the use of an oxide substrate. However, the atomic concentration of Ca was much higher than Zn and N compared with the case of $CaZn_2N_2$ on GaN. Therefore, we conclude that *p*-type conduction in $CaZn_2N_2$ on YSZ derived from nonstoichiometry, where Zn and N vacancies supplied holes and electrons, respectively. Although, we attribute the supply of both p- and n-type carriers to carrier compensation, Zn vacancies should be dominant in this case because the atomic concentration of Zn is less than that of N over the whole film. These results suggest that if O contamination is suppressed and the formula stoichiometry is realized, the promising expected and predicted optical and electronic characteristics of $CaZn_2N_2$ should be realized.

**Conclusions**

In summary, we have successfully grown epitaxial thin films of a ternary Zn-based nitride $CaZn_2N_2$ semiconductor by RF-MBE, which has previously only been stabilized in the polycrystalline bulk form at an ultrahigh pressure at 5 GPa. The key conditions were to individually and precisely tune $P_{Zn}$ and $P_{Ca}$ and use GaN template layers and an active N-radical source in bright mode, generated at an rf power >300 W. Because *in situ* PLD and R-SPE did not create a $CaZn_2N_2$ phase, this RF-MBE technique represents a promising way to stabilize $CaZn_2N_2$ epitaxial films. Other rf-plasma sources such as rf sputtering methods are also effective for stabilizing $CaZn_2N_2$ and related nitride epitaxial films, as demonstrated by recent reports of the stabilization of new metastable nitride phases such as $ZnSnN_2$ [13] and $Cu_3N$ [14, 15]. Additionally, this first demonstration of epitaxial growth of $CaZn_2N_2$ can be widely and easily extended to other related alloying nitrides such as $Ca(Mg_{1-x}Zn_x)_2N_2$ (3.3–1.9 eV) [2, 3], $(Sr_{1-x}Ca_x)Zn_2N_2$ (1.6–1.9 eV) [2], and $Ca(Cd_{1-x}Zn_x)_2N_2$ (1.9–0.4 eV) [2], leading to wide band gap control from the ultraviolet to infrared regions only in one isomorphic nitride system. The achieved optical and electrical properties remain poorer than those predicted. However, further optimization of the processing parameters to increase the nitrogen chemical potential together with growth at temperatures higher than 250 °C should improve the crystal quality of this promising ternary nitride semiconductor.

**Acknowledgments**

This work was supported by JST CREST Grant JPMJCR17J2, Japan. The authors thank Yuji Kondo and Michiko Sato for performing EPMA and FE-AES measurements, respectively. H. Ho. was supported by the Ministry of Education, Culture, Sports, Science, and Technology (MEXT) through the Element Strategy Initiative to Form Core Research Center. H. Hi. was also supported by the Japan Society for the Promotion of Science (JSPS) through Grants-in-Aid for Scientific Research (A) and (B) (Grants 17H01318 and 18H01700) and Support for Tokyotech Advanced Research (STAR).

**Supporting Information**

The Supporting Information is available on the ACS website.

**References**

1. Auf der Maur, M.; Pecchia, A.; Penazzi, G.; Rodrigues, W.; Di Carlo, A. Efficiency Drop in Green InGaN/GaN Light Emitting Diodes: The Role of Random Alloy Fluctuations. *Phys. Rev. Lett.* **2016**, 116, 027401.
2. Hinuma, Y.; Hatakeyama, T.; Kumagai, Y.; Burton, L. A.; Sato, H.; Muraba, Y.; Iimura, S.; Hiramatsu, H.; Tanaka, I.; Hosono, H.; Oba, F. Discovery of earth-abundant nitride semiconductors by computational screening and high-pressure synthesis. *Nat. Commun.* **2016**, 7, 11962.
3. Tsuji, M.; Hiramatsu, H.; Hosono, H. Tunable light-emission through the range 1.8–3.2 eV and p-type conductivity at room temperature for nitride semiconductors, Ca(Mg$_{1-x}$Zn$_x$)$_2$N$_2$ ($x = 0 - 1$). Submitted to *Inorg. Chem.* arXiv:1902.10495 https://arxiv.org/abs/1902.10495
4. Ohta, H.; Nomura, K.; Orita, M.; Hirano, M.; Ueda, K.; Suzuki, T.; Ikuhara, Y.; Hosono, H. Single-Crystalline Films of the Homologous Series InGaO$_3$(ZnO)$_m$ Grown by Reactive Solid-Phase Epitaxy. *Adv. Funct. Mater.* **2003**, 13, 139 – 144.
5. Hiramatsu, H.; Ueda, K.; Ohta, H.; Orita, M.; Hirano, M.; Hosono, H. Heteroepitaxial growth of a wide-gap p-type semiconductor, LaCuOS. *Appl. Phys. Lett.* **2002**, 81, 598 – 600.
6. Park, C.-H.; Keszler, D. A.; Yanagi, H.; Tate, J. Gap modulation in MCu[Q$_{1-x}$Q'$_x$]F (M = Ba, Sr; Q, Q' = S, Se, Te) and related materials. *Thin Solid Films* **2003**, 445, 288 – 293.
7. Nomura, K.; Ohta, H.; Ueda, K.; Kamiya, T.; Orita, M.; Hirano, M.; Suzuki, T.; Honjyo, C.; Ikuhara, Y.; Hosono, H. Growth mechanism for single-crystalline thin film of InGaO$_3$(ZnO)$_5$ by reactive solid-phase epitaxy. *J. Appl. Phys.* **2004**, 95, 5532 – 5539.
8. Hiramatsu, H.; Ohta, H.; Suzuki, T.; Honjo, C.; Ikuhara, Y.; Ueda, K.; Kamiya, T.; Hirano, M.; Hosono, H. Mechanism for Heteroepitaxial Growth of Transparent P-Type Semiconductor: LaCuOS by Reactive Solid-Phase Epitaxy. *Cryst. Growth Des.* **2004**, 4, 301 – 307.
9. Fox, M. "Optical Properties of Solids" **2010**, Oxford University Press, second edition, United States.
10. Ptak, A. J.; Holbert, L. J.; Ting, L.; Swartz, C. H.; Moldovan, M.; Giles, N. C.; Myers, T. H.; Van Lierde, P.; Tian, C.; Hockett, R. A.; Mitha, S.; Wickenden, A. E.; Koleske, D. D.; Henry, R. L. Controlled oxygen doping of GaN using plasma assisted molecular-beam epitaxy. *Appl. Phys. Lett.* **2001**, 79, 2740 – 2472.
11. Slack, G. A.; Schowalter, L. J.; Morelli, D.; Freitas Jr., J. A. Some effects of oxygen impurities on AlN and GaN. *J. Cryst. Growth* **2002**, 246, 287 – 298.
12. Ohtomo, A.; Tamura, K.; Saikusa, K.; Takahashi, K.; Makino, T.; Segawa, Y.; Koinuma, H.; Kawasaki, M. Single crystalline ZnO films grown on lattice-matched ScAlMgO$_4$(0001) substrates. *Appl. Phys. Lett.* **1999**, 75, 2635 – 2637.
13. Lahourcade, L.; Coronel, N. C.; Delaney, K. T.; Shukla, S. K.; Spaldin, N. A.; Atwater, H. A. Structural and Optoelectronic Characterization of RF Sputtered ZnSnN$_2$. *Adv. Mater.* **2013**, 25, 2562–2566.
14. Caskey, C. M.; Richards, R. M.; Ginley, D. S.; Zakutayev, A. Thin film synthesis and properties of copper nitride, a metastable semiconductor. *Mater. Horiz.* **2014**, 1, 424 – 430.
15. Matsuzaki, K.; Harada, K.; Kumagai, Y.; Koshiya, S.; Kimoto, K.; Ueda, S.; Sasase, M.; Maeda, A.; Susaki, T.; Kitano, M.; Oba, F.; Hosono, H. High-Mobility p-Type and n-Type Copper Nitride Semiconductors by Direct Nitriding Synthesis and In Silico Doping Design. *Adv. Mater.* **2018**, 30, 1801968.

Supporting Information for "Heteroepitaxial Thin-Film Growth of a Ternary Nitride Semiconductor CaZn$_2$N$_2$"

Masatake Tsuji,[1] Kota Hanzawa,[1] Hiroyuki Kinjo,[1] Hidenori Hiramatsu,[1,2,*] and Hideo Hosono [2]

1: Laboratory for Materials and Structures, Institute of Innovative Research, Tokyo Institute of Technology, Mailbox R3-3, 4259 Nagatsuta-cho, Midori-ku, Yokohama 226-8503, Japan

2: Materials Research Center for Element Strategy, Tokyo Institute of Technology, Mailbox SE-1, 4259 Nagatsuta-cho, Midori-ku, Yokohama 226-8503, Japan

*) E-mail: h-hirama@mces.titech.ac.jp

**Text S1**

The CaZn$_2$N$_2$ synthesized via a solid-state reaction between binary precursors of Ca$_3$N$_2$ and Zn$_3$N$_2$ under a high pressure of 5 GPa contains a large amount of metallic Zn impurities [2]. Hence, for in-situ PLD growth, we initially removed the metallic-Zn impurity to produce highly-purified CaZn$_2$N$_2$ polycrystalline bulk PLD targets. **Figure S1a** shows the purification procedure used to prepare the Zn-removed CaZn$_2$N$_2$ by a chemical reaction between Zn metal and I$_2$. Under heating the as-prepared CaZn$_2$N$_2$ powder containing metallic Zn with I$_2$ in Ar at 150 °C for 10 min, the Zn metal impurity selectively reacted with I$_2$, resulting in the formation of a binary compound of ZnI$_2$. Because ZnI$_2$ can easily dissolve in diethyl ether solution and subsequently decompose to Zn$^{2+}$ and I$^-$ ions, we added solvent to the CaZn$_2$N$_2$ heated with I$_2$, removed the solvent, and dried the resulting powder. The XRD pattern of the obtained powder is shown in **Figure S1b**. From the Rietveld refinement, we found that the atomic concentration of the Zn metal impurity was decreased from 31 mol% [2] to 0.7 mol% through the I$_2$ treatment. The purified CaZn$_2$N$_2$ powder was pressed at 200 MPa under a cold isostatic pressure to form disk-shaped targets for in-situ PLD growth.

**Figure S2a** shows the $T_s$ dependence of XRD patterns of ~450 nm-thick films deposited by in-situ PLD under a N$_2$-gas atmosphere of ~5 Pa on silica glass and YSZ (111) as substrates at $T_s$ = room temperature (RT) and 800 °C, respectively, where the growth rate was fixed at ~15 nm/min. The obtained crystalline phase was only CaO, where oxygen derived mainly from oxide impurities in the PLD bulk targets and the YSZ substrate. Moreover, EPMA results (**Figure S2b**) indicated that the atomic ratios of the films were far from the stoichiometry; the atomic ratios of [Zn]/[Ca] were 1.20 and 0.03, and those of [N]/[Ca] were 0.16 and 0.005 for the films grown at $T_s$ = RT and 800 °C, respectively. On the basis of these results, we speculated that optimization of the chemical composition should stabilize the CaZn$_2$N$_2$ phase and tried to grow the CaZn$_2$N$_2$ by post-deposition thermal annealing with the use of R-SPE, as shown in **Figure S2c**. In the process, a buffer layer of Zn metal or Zn$_3$N$_2$ was introduced between the amorphous Ca-Zn-N, deposited at RT, and the YSZ substrate to achieve a stoichiometric chemical composition, in which Zn$_3$N$_2$ was grown under a ~5 Pa RF-generated active N-radical gas at 100 W. The multi-layered samples were capped with YSZ plates, wrapped by Pt wire to attach the sample to plate, and vacuum-sealed in silica-glass ampules with Zn$_3$N$_2$ powder to achieve high chemical potentials of Zn and N. **Figure S2d** shows XRD patterns of the sample based on a Zn-buffer layer before and after the thermal annealing at annealing temperatures ($T_a$) of 300 and 500 °C, where Zn and amorphous Ca-Zn-N layers were deposited to be ~20 and ~180 nm; i.e., $t$ = ~200 nm in total. Before the thermal treatment, only diffraction features originating from Zn were detected; conversely, we observed segregation of ZnO and Zn after annealing, where the oxygen derived from non-crystalline phases in the film such as CaO, $Ca(OH)_2$, and $CaO \cdot x(H_2O)$, which formed from CaO through air exposure. Although an unambiguous $CaZn_2N_2$ phase was not achieved, the atomic ratio of [Zn]/[Ca] reached 1.5 (**Figure S2e**) through annealing at $T_a$ = 300 °C. This value is closer to the formal stoichiometry than the results of as-prepared samples ([Zn]/[Ca] = 1.4) or the sample annealed at $T_a$ = 500 °C (1.4). To achieve a much higher chemical pressure of N and prevent nucleation of the oxide, we prepared samples with a ~10 nm-thick $Zn_3N_2$-buffer layer and ~90 nm-thick amorphous Ca-Zn-N (entire $t$ = 100 nm) deposited at RT. **Figure S2f** shows the results of the post-deposition thermal annealing, in which we selected $T_a$ = 300 °C, based on previous results with the use of the metal Zn-buffer layer. Although the thermal annealing led to a stoichiometric atomic ratio ([Zn]/[Ca] = 2.0), as evaluated by X-ray fluorescence, crystalline $CaZn_2N_2$ was not nucleated; the $Z_3N_2$ remained similar to the as-prepared sample.

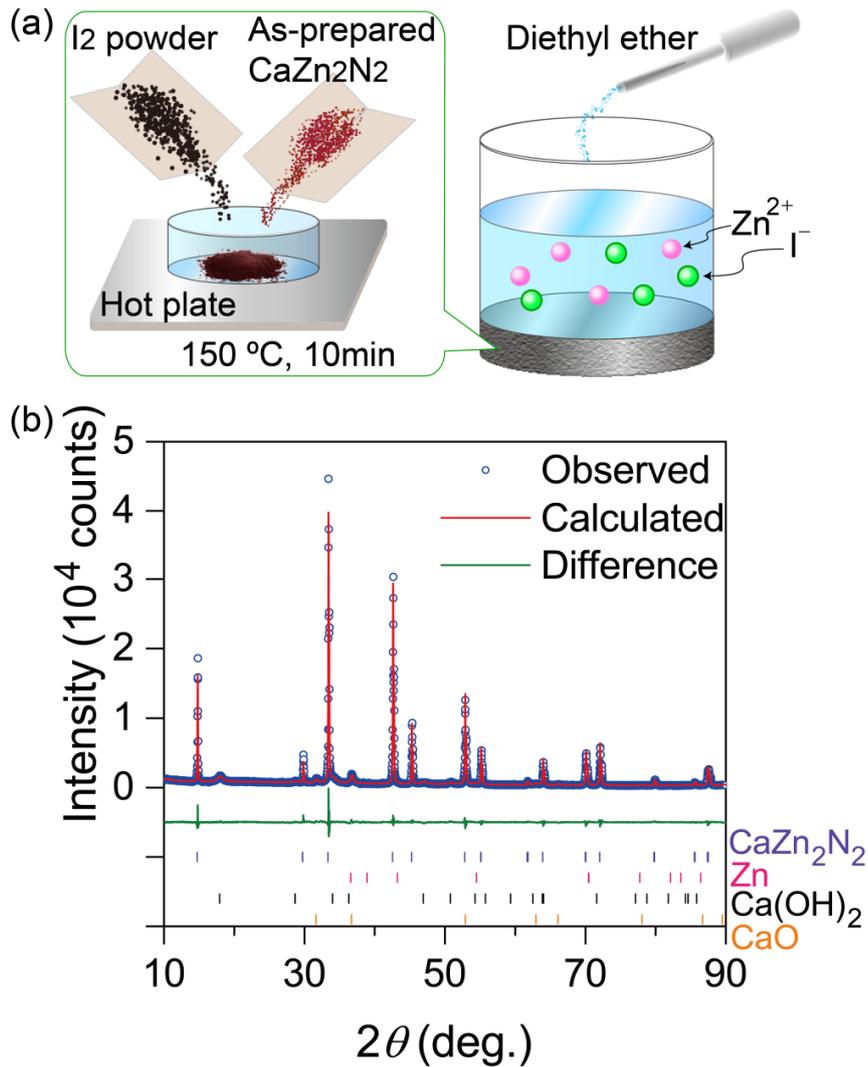

**Figure S1.** Preparation of impurity-Zn-removed $CaZn_2N_2$ polycrystalline bulk as a PLD target disk. (a) Procedure to remove Zn metal from the as-prepared $CaZn_2N_2$ powder. The as-prepared $CaZn_2N_2$ was mixed with $I_2$ at 150 °C for 10 min, to enable the chemical reaction: $Zn + I_2 \rightarrow ZnI_2$. The resulting mixture was stirred in diethyl ether solution. In this process, $Zn^{2+}$ and $I^-$ decomposed from $ZnI_2$ dissolved in the solvent. By drying and evaporating the solvent, purified $CaZn_2N_2$ was obtained and then pressed under a cold isostatic pressure of 200 MPa to form a pellet. (b) XRD pattern of the resulting Zn-removed $CaZn_2N_2$ powder. Concentrations of $CaZn_2N_2$, Zn, $Ca(OH)_2$, and CaO were respectively 45.4, 0.7, 32.6, and 21.3 mol%, as quantified by Rietveld analysis.

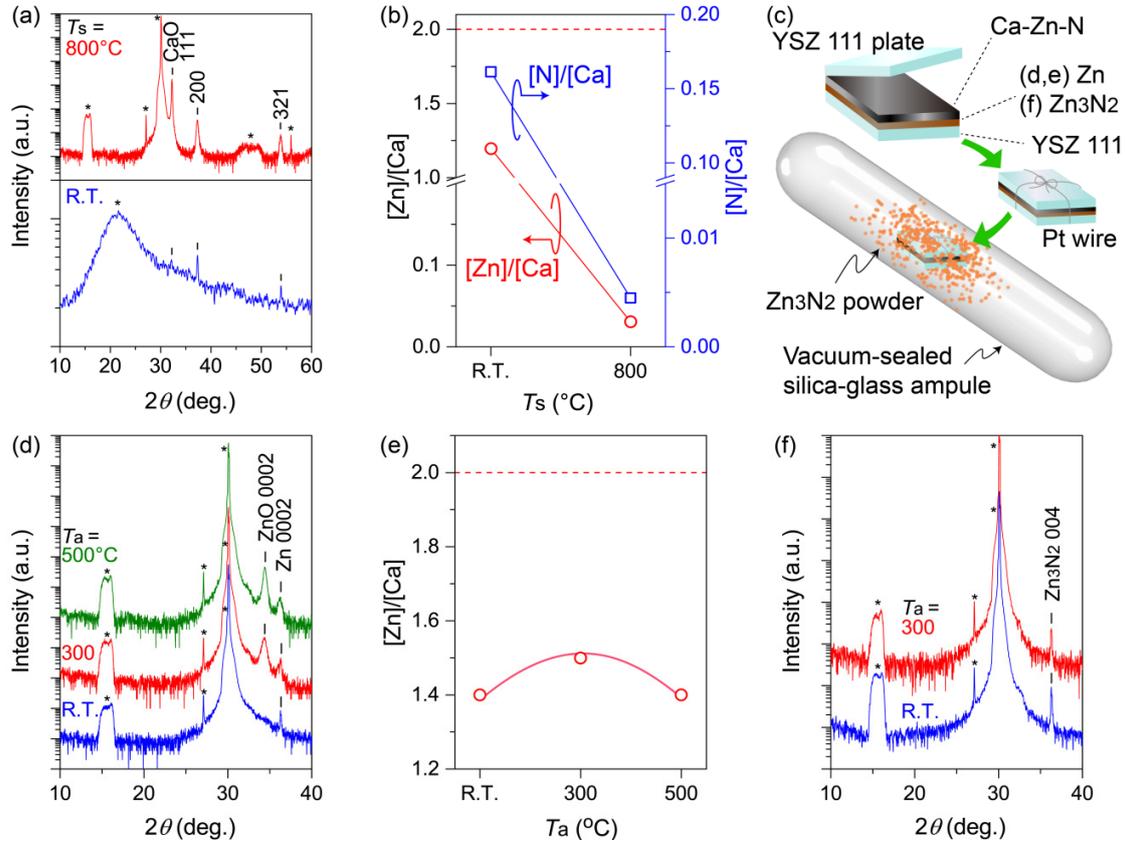

**Figure S2.** Results of attempts to fabricate CaZn$_2$N$_2$ thin films with the use of in-situ PLD and the post-deposition thermal annealing process by a R-SPE technique. (a) XRD patterns of thin films grown from Zn-removed CaZn$_2$N$_2$ by in-situ PLD under $P_N$ = ~5 Pa at $T_s$ = 800 °C on YSZ (top) and room temperature (RT) on silica glass (bottom). (b) Chemical compositional analysis of the thin films in (a) as evaluated by EPMA. Red circles and blue squares indicate atomic ratios of Zn and Ca ([Zn]/[Ca]), and those of N and Ca ([N]/[Ca]), respectively. (c) Post-deposition thermal annealing procedure based on the R-SPE technique. Amorphous Ca-Zn-N films were grown at RT on Zn or Zn$_3$N$_2$/YSZ deposited under a vacuum and rf-generated N-radical gas at 100 W, respectively. These films were capped with a YSZ plate and wrapped with Pt wire. Subsequently, the samples were thermally annealed at $T_a$ = 300 or 500 °C in vacuum-sealed silica-glass ampules. (d) XRD patterns of the films obtained through the post-deposition thermal annealing with amorphous Ca-Zn-N/Zn/YSZ configuration. (e) Chemical composition as analyzed by X-ray fluorescence at each $T_a$. (f) XRD patterns before and after the thermal annealing of amorphous Ca-Zn-N/Zn$_3$N$_2$/YSZ at $T_a$ = 300 °C. Asterisks denote diffraction features from the substrates.

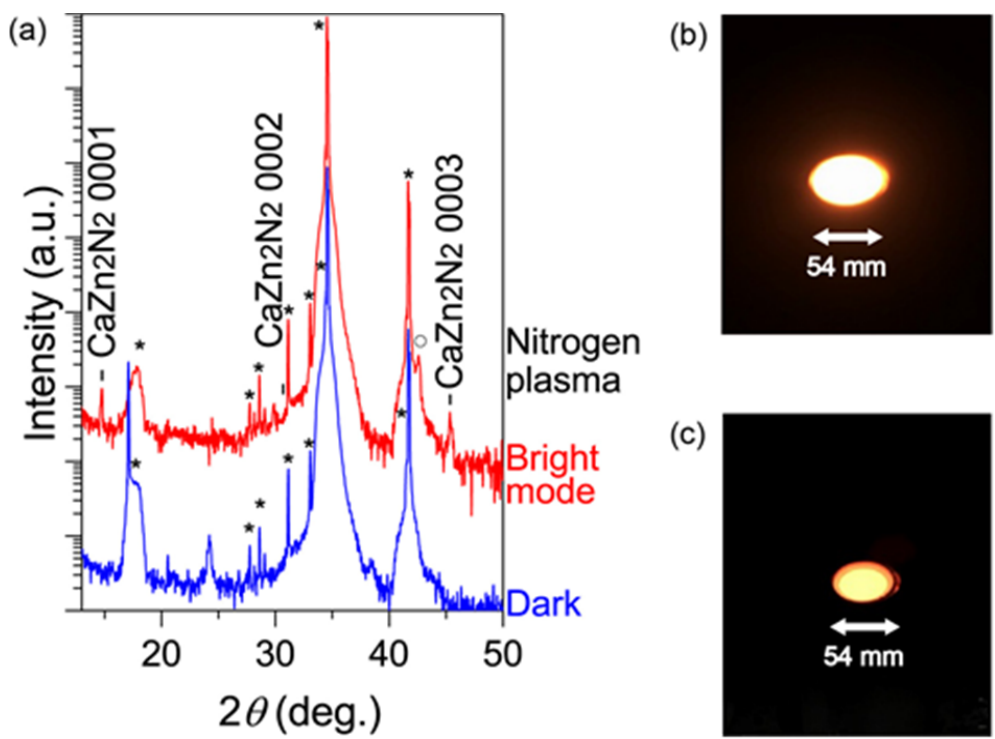

**Figure S3.** Influence of N plasma with bright (> 300 W, top) and dark (< 300 W, bottom) modes for stabilization of $CaZn_2N_2$ phase on GaN template layers. (a) XRD patterns of the thin films grown with different nitrogen plasma. Circles denote diffraction features from possible impurities of $CaZn_5$ and squares indicate those of $CaZn_2$ and $CaZn_3$. Asterisks indicate diffraction peaks from the substrate. Photographs of rf generator with (b) Bright mode and (c) Dark mode.

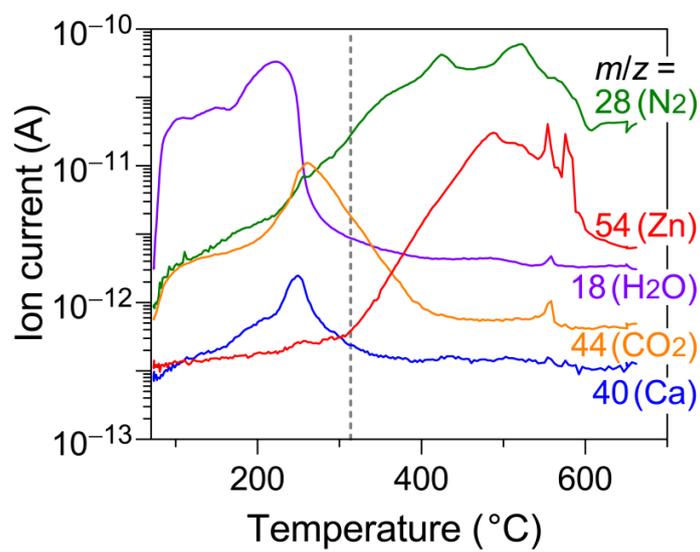

**Figure S4.** Thermal desorption spectra of polycrystalline $CaZn_2N_2$. Dashed vertical line at 320 °C indicates temperature for the start of decomposition; i.e., when the major constituent Zn in $CaZn_2N_2$ starts to evaporate.

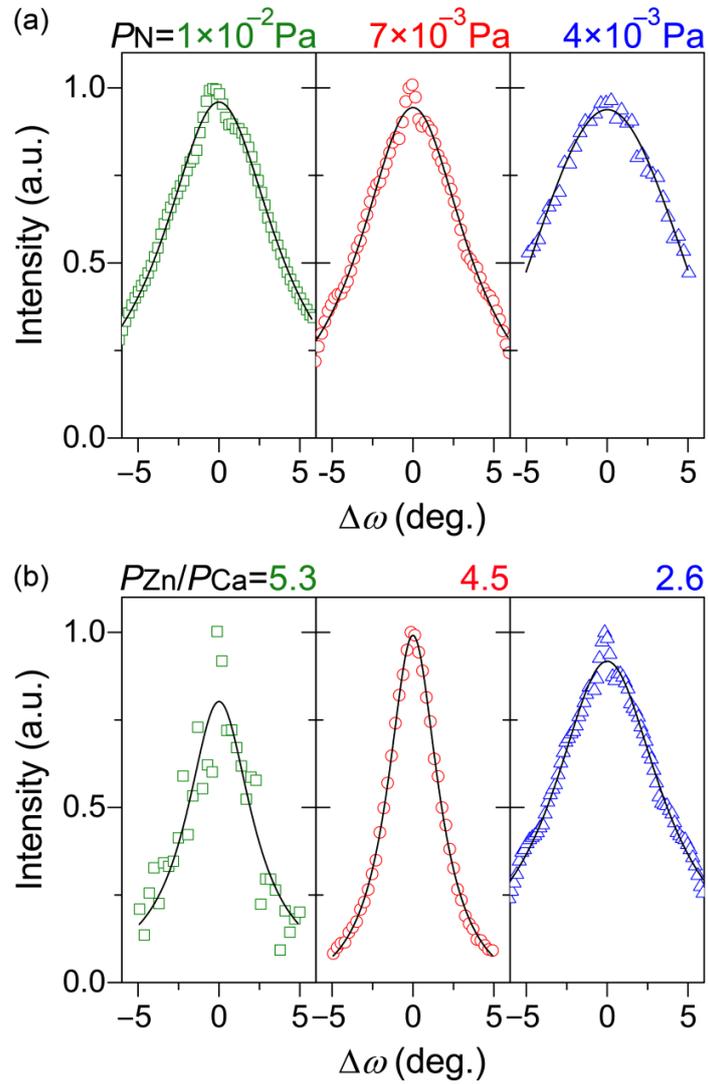

**Figure S5.** Out-of-plane rocking curves of the 0001 diffraction features in $CaZn_2N_2$ films on GaN grown at (a) different $P_N$ and (b) $P_{Zn}/P_{Ca}$. Solid lines in each curve are the results of least-square fitting.